# Artificial Intelligence and the Future of Psychiatry:

# Qualitative Findings from a Global Physician Survey

**Authors and affiliations:**

[1,2]C. Blease, [3,4]C. Locher, [5]M. Leon-Carlyle, [6]P. M. Doraiswamy

[1]Program in Placebo Studies, General Medicine and Primary Care, Beth Israel Deaconess Medical Center/Harvard Medical School, Boston, USA

[2]School of Psychology, University College Dublin, Ireland

[3] Division of Clinical Psychology and Psychotherapy, University of Basel, Switzerland

[4]University of Plymouth, UK

[5]University of Toronto, Canada

[6]Departments of Psychiatry and Behavioral Science, and Medicine, Duke University Medical School, Durham, North Carolina, USA.

**Address for correspondence:**

C. Blease

General Medicine and Primary Care

Beth Israel Deaconess Medical Center

330 Brookline Avenue

Boston 02215

cblease@bidmc.harvard.edu

**Abstract**

*Background*: The potential for machine learning to disrupt the medical profession is the subject of ongoing debate within biomedical informatics.

*Objective*: This study aimed to explore psychiatrists' opinions about the potential impact innovations in artificial intelligence and machine learning on psychiatric practice

*Methods*: In Spring 2019, we conducted a web-based survey of 791 psychiatrists from 22 countries worldwide. The survey measured opinions about the likelihood future technology would fully replace physicians in performing ten key psychiatric tasks. This study involved qualitative descriptive analysis of written responses ("comments") to three open-ended questions in the survey.

*Results*: Comments were classified into four major categories in relation to the impact of future technology on: (1) patient-psychiatrist interactions; (2) the quality of patient medical care; (3) the profession of psychiatry; and (4) health systems. Overwhelmingly, psychiatrists were skeptical that technology could replace human empathy. Many predicted that 'man and machine' would increasingly collaborate in undertaking clinical decisions, with mixed opinions about the benefits and harms of such an arrangement. Participants were optimistic that technology might improve efficiencies and access to care, and reduce costs. Ethical and regulatory considerations received limited attention.

*Conclusions*: This study presents timely information on psychiatrists' views about the scope of artificial intelligence and machine learning on psychiatric practice. Psychiatrists expressed divergent views about the value and impact of future technology with worrying omissions about practice guidelines, and ethical and regulatory issues.

# Introduction

## Background

Worldwide it is estimated that 1 in 6 people suffer from mental health disorders, and the personal and economic fallout is immense.[] Psychiatric illnesses are among the leading causes of morbidity and mortality; by 2030 this burden is estimated to cost the global economy $16 trillion. [] Among younger people, suicide is the second or third leading cause of death.[] Older generations are also affected by mental illness: currently, an estimated 50 million people suffer from dementia worldwide, and the World Health Organization (WHO) predicts this will rise to 80 million by 2030.[] Stigmatization, low funding and lack of resources – including considerable shortages of mental health professionals – pose significant barriers to psychiatric care.[] According to recent WHO data, discrepancies in per-capita availability of psychiatrists is 100 times lower than in affluent countries.[] Indeed, even in wealthy countries, such as the USA – which has around 28,000 psychiatrists[] – those living in rural or poverty-stricken urban communities experience inferior access to adequate mental health care. It is anticipated that demographic and societal changes will put even greater pressure on mental health resources in the forthcoming decades.[] These pressures include: ageing populations; increased urbanization (with associated problems of overcrowding, polluted living conditions, higher levels of violence, illicit drugs, and lower levels of social support); migration, at the highest rate recorded in human history; and the use of electronic communications which has amplified concerns about the effects of the internet on mental health and sociality.[]

Against these myriad challenges, recent debate has centered on the potential of big data, machine learning (ML) and artificial intelligence (AI) to revolutionize the delivery of healthcare.[] Some medical informaticians argue that the core functions of physicians – gathering and monitoring patient information, diagnostics, prognostics, and formulating personal treatment plans are vulnerable to disintermediation.[] Other AI experts predict that, in the future, physicians will always play a role in medical care with 'man and machine' working as 'team-players'.[] Aside from forecasts by medical professionals and informaticians, there is increasing interest among patients to use mobile technologies to monitor their health. For example, a recent US survey reported that 70 per cent of patients had an interest in using mobile technologies to monitor their mental health.[]

**Objectives**

Amid the debate, hype, and uncertainties about the impact of AI on the future of medicine, limited attention has been paid to the views of practicing clinicians including psychiatrists[] – though in the last year there is evidence that this changing.[] To address this research gap, we employed quantitative methods to investigate the global psychiatric community's opinions about the potential impact of future technologies to replace key physician tasks in mental health care. Given the potential for more nuanced insights acquired through qualitative methods, we incorporated 3 open-ended questions into the survey (see Table 1). Our aim was to provide a preliminary investigation into psychiatrists' views on the themes of the quantitative online survey: namely, participants' perspectives on the influence (if any) of AI/ML on the profession, including potential benefits and harms to psychiatric practice. To our knowledge, this is the first qualitative investigation of psychiatrists' opinions about AI/ML on the future of the profession.

**Methods**

**Main Survey**

A complete description of the survey methods and quantitative results has been published previously.[] In summary, we conducted an anonymous global Web-based survey of psychiatrists registered with Sermo, a secure digital online social networking for physicians, and for conducting survey research.[] Participants were randomly sampled from membership of the Sermo.org []. This is the one of the largest online medical networks in the world, with 800,000 users from 150 countries across Europe, North and South America, Africa, and Asia, employed in 96 medical specialties. Users are registered and licensed physicians. Invitations were emailed and displayed on the Sermo.org home pages of randomly selected psychiatrists in May 2019, with quasi-stratification: the aim was to recruit one third of participants from the USA, one third from Europe, and one third from the rest of the world. As this was an exploratory study, we aimed to target a sample size of roughly 750 participants to approximate a previous survey of general practitioners' views, on which the current project was based.[] The survey was closed with 791 respondents. This was an anonymous survey and an analysis of de-identified survey data was deemed exempt research by Duke University Institutional Review Board. Invited participants were advised that their identity would not be disclosed to the research team, and all respondents gave informed consent before participating.

The study team devised an original survey instrument specifically designed to investigate psychiatrists' opinions about the impact of future technologies on primary care (see Multimedia Appendix 1). We avoided terms such as "algorithms" in favor of generic descriptors such as "machines" and "future technology." This was in part to avoid any confusion among physicians unfamiliar with this terminology and to avert technical debates about the explanatory adequacy of specific AI terms of art. The survey was developed in consultation with psychiatrists in the USA (n=2) and was pretested with psychiatrists from other countries (n=9) to ensure face validity.

**Qualitative Component**

To maximize response rate for the qualitative component, the survey instrument included three open-ended questions that allowed participants to respond in more detail on the topic of the questionnaire (see Table 1). Comments not in English were translated by Sermo; this process was undertaken by experienced medical text translators, subject to further proofreading and in-house checks. Descriptive content analysis was used to investigate these responses.[] Responses were collated and imported into QCAmap (coUnity Software Development GmbH) for analysis. The comment transcripts were initially read numerous times by CB, CL, and MLC to achieve familiarization with the participant responses. Afterward, an inductive coding process was employed. This widely used method is considered an efficient methodology for qualitative data.[] A multistage analytic process was conducted: First, we defined the three open-ended questions as our main research questions. Second, we worked through the responses line by line. Brief descriptive labels ("codes") were applied to each comment. Multiple codes were applied to comments with multiple meanings. Comments and codes were reviewed by CB, CL and MLC. Third, after working through a significant amount of text, CB, CL and MLC met to discuss coding decisions, and subsequent revisions were made. This process led to a refinement of codes. Finally, first-order codes were grouped into second-order categories based on the commonality of their meaning to provide a descriptive summary of the responses.[] We followed the rules of summarizing qualitative content analysis for this step.[]

# Results

## Overview

As outlined in the quantitative survey, 791 psychiatrists responded from 22 countries representing North America, South America, Europe, and Asia-Pacific.[] Of the participants, 70% were male; and 61% were aged 45 or older (see Table 2). All respondents left comments (26,470 words) which were typically brief (1 phrase or 2 sentences).

As a result of the iterative process of content analysis, four major categories were identified in relation to the impact of future technology on: (1) patient-psychiatrist interactions; (2) the quality of patient medical care; (3) the profession of psychiatry; and (4) health systems. These categories were further subdivided into themes, which are described below with illustrative comments; numbers in parentheses are identifiers ascribing comments to individual participants.

## Impact of Future Technology on Patient-Psychiatrist Interactions

A foremost concern of future technology on psychiatry was the perceived "*loss of empathy*", and absence of a therapeutic interpersonal relationship in the treatment of mental health patients.

### *Empathy*

Numerous comments reflected considerable skepticism that future technology could provide empathic care; an underlying assumption was this was necessarily a human capacity. Some participants were adamant about this; for example:

> *It is a total illusion. A robot is incapable of being empathetic in a caring relationship.* [Participant 259]

> *Psychiatry is incompetent and incomplete without empathy. I doubt a machine could ever empathize with a live human being... I don't think affect of patient and mood, feelings, emotions can be analyzed accurately.* [Participant 61]

> *Lack of empathy will be huge.* [Participant 397]

Although most responses were short – for example, "*no empathy*", or "*lack of empathy and humanity*" – a significant number of respondents also perceived limitations that technology could ever accurately detect human emotions via verbal or nonverbal cues; for example,

> *AI cannot properly assess the human side of the problems including thing as basic as facial expression and posture, eye contact, etc*. [Participant 80]

> *AI could be overwhelmed as it tries to sort out body language, affect, lying, and conversational subtleties.* [Participant 462]

> *It won't be able to read the subtext of conversation.* [Participant 404]

Only a small minority of comments hinted at the benefits of machine technology in augmenting empathetic care including the detection of emotions:

> *Objectively (without countertransference) giving weight to micro-facial recognition/paired with some sort of emotional detection recognition software.* [Participant 53]

> *May ask questions in a non-judgmental way – likely to emulate empathy*. [Participant 548]

***The therapeutic relationship***

Another dominant view was the broader implications of technology for the therapeutic relationship with the majority of comments anticipating communication problems, lack of rapport, and the potential harms to patients. Notably, the majority of responses assumed that future technology would incur loss of contact with clinicians and even incur harm. For example:

> *Complete lack of human interactions central to psychiatry services!* [Participant 35]

> *Absence of human relationships that are so vital for healing.* [Participant 438]

> *Dehumanizing: patients could feel invalidated and ignored, loss of therapeutic relationship.* [Participant 35]

> *It will increase psychiatric illness. Clients will be more isolated and seek personal interaction.* [Participant 433]

Some respondents indicated that patients would prefer to seek help from humans; for example:

> *Patients are still going to want to talk to an actual human being about deep seated thoughts, and feelings.* [Participant 38]

> *Patients will always want the human connection and if not from psychiatrists than shamans, natural healers, or other alternative practitioners.* [Participant 534]

Taking an opposing view, a few psychiatrists suggested that future technology might improve on human interactions; for example:

> *People interacting with machines is much easier than with fellow human beings. We are assessing this phenomenon today when children are having "best friends" who they have only met through Facebook. It is very comfortable to have an "avatar" as a friend. Because, we select when to cut them off.* [Participant 413]

> *Would do a better job than our autism and ADHD service most staff [of] which are empathic as a robot anyway.* [Participant 782]

> *When I accept patients that have seen other providers in my town, I am ever amazed and disappointed in the report of the care they've received. Seems patients don't seem to connect with doctors (or any provider for that matter) any more. So if there is no interpersonal connection/relationship, why not type into a computer?* [Participant 95]

*Telepsychiatry*

Only a few participants predicted an increase in the use of telepsychiatry including the use of "*psychotherapy via Skype*". Notably, these comments tended to be neutral with respect to the potential benefits or harms of telepsychiatry on doctor-patient interactions; for example:

> *Telepsychiatry will prevail.* [Participant 458]

> *Telemedicine – face-to-face contact no longer required.* [Participant 602]

*Trust, privacy, and confidentiality*

Similarly, implications for the fiduciary doctor-patient relationship also received very limited attention. However, some comments suggested that patients would not find technology acceptable in their care leading to lower rates of satisfaction, resistance, or even refusal to be treated. For example:

> *Potential harms include alienating or breeding distrust in patients.* [Participant 364]

> *I can see trust as an issue, in someone who knows they're interacting with a computer which could then block or greatly limit potential benefits.* [Participant 771]

> *The great disadvantage of impersonal treatment via an app will only convince a few patients.* [Participant 149]

> *I suspect that there may be feelings of antipathy towards AI/tech due to job displacement which would not make for very therapeutic interactions.* [Participant 364]

For others, trustful interactions could be vulnerable to exploitation or manipulation from patients, including faking illnesses; for example:

> *If a patient simulates a disease AI might not be able to determine it.* [Participant 124]

> *Machines could be manipulated by sociopaths.* [Participant 548]

However, one psychiatrist took an opposing and more optimistic view, responding that patients may exhibit greater trust in technology than in clinicians:

> *People will have the confidence in bold technology as they'll feel more confident that they can be treated more safely.* [Participant 779]

The topic of data safety, misuse of data, and questions of privacy, received only a small number of truncated comments; for example:

> *Can't keep patients' privacy – the data will be hacked.* [Participant 758]

> *My only worry would be data sharing arrangements.* [Participant 62]

Only one participant suggested that mental health patients may be at greater risk of harm from loss of confidentiality with new technologies:

> *Given how vulnerable mentally ill patients are, AI must be monitored carefully. Already senior citizens fall for financial scams and predators targeting them. It would be important to protect confidentiality.* [Participant 79]

**Impact of Future Technology on the Quality of Patient Medical Care**

Implications of future technology for patient care received considerable attention, and a mixture of opinions were offered about potential benefits and harms.

*Medical error*

Many respondents suggested that technology would reduce errors or improve accuracy in clinical decisions – including in diagnostics and treatment decisions. For example:

> *Eliminate the human error.* [Participant 50]

> *Will speed up diagnosis and assess patient's condition more accurately.* [Participant 583]

> *The benefits would be greater reliability in diagnosis and prognosis, being able to choose specific customized treatment plans after analysis.* [Participant 562]

A few comments suggested that technology could improve care by identifying drug-drug interactions or potential contraindications to treatment; for example:

> *I think it can help identify safer prescribing/dangerous combinations, (…) data analysis of risk factors to recommend treatment setting.* [Participant 359]

> *Watching drug interactions and appropriate options of meds.* [Participant 63]

More broadly, a minority of comments were very enthusiastic about the role of technology in patient care; for example:

> *Only benefits: improvement in the care of the patient. I don't see any harm.* [Participant 155]

> *Only improvements!* [Participant 233]

> *Clearer approach, no possible damage.* [Participant 539]

In contrast to these optimistic responses, however, a considerable number of comments suggested that future technology would lead to an increase in medical error. Many of these comments specifically referred to an increased risk of diagnostic error; for example:

> *Damage: the wrong diagnoses, false positives, false negatives.* [Participant 129]

> *Automated assessments with limited and strict protocols may limit flexibility of treatment and diagnosis for complex cases.* [Participant 65]

> *This will lead to many misdiagnoses and inappropriate medications.* [Participant 419]

Going further, some respondents were adamant about the lack of potential benefits of technology; for example:

> *I don't foresee any benefits at all.* [Participant 713]

> *Give a computer an ego and it will kill us all.* [Participant 491]

> *I cannot think of any possible benefits.* [Participant 589]

Finally, opposing these polarized perspectives, some psychiatrists admitted that they were unfamiliar with the topic of artificial intelligence, and refrained from taking a position; for example:

> *I am curious about the role of artificial intelligence.* [Participant 573]
>
> *I would like to know more about this topic.* [Participant 186].
>
> *I don't know and I wouldn't know what to say about it… I have to get more familiar with the subject.* [Participant 629]

*Avoiding bias in clinical judgments*

Many participants anticipated that artificial intelligence would be "*more objective*", "*fairer*", or "*unbiased*" compared to human psychiatrists; for example:

> *Less bias due to race or gender.* [Participant 64]
>
> *Computers can be objective and are not subject to recent recall of humans.* [Participant 400]
>
> *The advantage of AI is that a situation can be assessed/weighted without any personal bias.* [Participant 710]
>
> *Reduction in number of errors caused by emotional factors.* [Participant 212]

*Improved detection and monitoring of mental health*

A number of respondents commented on the possibility for improved preventive mental health including earlier diagnosis and increased screening

> *Early diagnosis advantage and therefore reduced the current slow diagnosis of certain diseases, improvement in prognosis – real prevention.* [Participant 732]

> *Possibility of monitoring over a prolonged period, especially with repeated acute episodes that may be out of the practitioner's field.* [Participant 776]
>
> *We will also have the ability to utilize AI or machine learning in assessing individual patient profiles in whatever capacity that is attainable at that time to supplement or replace universal screenings.* [Participant 65]
>
> *Artificial intelligence will help us to calculate odds, to evaluate the risk of suicide and to detect when a patient has a high risk of decompensation.* [Participant 552]

Other respondents felt technology might facilitate the monitoring of treatment regimens; for example:

On the other hand, some comments were more doubtful that technology might aid preventive services; for example:

> *Lack of identification of at risk situations.* [Participant 188]
>
> *Suicidal tendency will be difficult to detect.* [Participant 123]
>
> *Inefficiencies in determining risk factors.* [Participant 693]

**Impact of Future Technology on the Profession**

Participants expressed a broad range of opinions about the impact of future technology on the profession: from outright replacement of psychiatrists to displacement of key functions of practice, and from skepticism about any change to uncertainty about the future. Responses also indicated a wide array of attitudes about the potential to influence of the field, from very negative to very positive with many psychiatrists displaying neutral perspectives.

*The status of the profession*

A common perspective was that specific aspects of the job would gradually be replaced by artificial intelligence with some psychiatrists predicting that this would lead to outright elimination; for example:

> *Jobs will reduce as AI will replace humans.* [Participant 15]

> *I believe psychiatrists (and physicians in general) will continue to be more and more marginalized by AI, and that more treatment decision making will be guided by AI in the future.* [Participant 528]

Some participants viewed change as a threat to the profession; for example:

> *More interference with the process of doctoring. See no good coming from it.* [Participant 27]

> *The end of the psychiatrist profession.* [Participant 144]

However, a few disagreed; for example:

> *Looking forward to artificial intelligence applications as quickly as possible.* [Participant 586].

*Facilitation of work activity*

Multiple comments predicted that future technology could facilitate the work of psychiatrists. Although most responses were rather short – for example, "*facilitation*", or "*it will make the job easy*"; lengthier responses included:

> *This help could enable the psychiatrist to carry on with his work and to be more effective.* [Participant 239]

> *It will help to relieve the burden on psychiatrists.* [Participant 268]

A considerable number of comments indicated psychiatrists will need to control and verify the technology-based results since machine recommendations would likely be error-prone; for example:

> *The problem is with being diagnosed by the machine. I think that the psychiatrist needs to verify the machine anyway. The machine cannot replace the human.* [Participant 267]

> *One potential harm is over reliance and not enough critical thinking about results, particularly results that support one's viewpoint.* [Participant 381]

Furthermore, multiple comments suggested that psychiatrists and future technology might have a "*job sharing*" arrangement with machines and humans complementing and enriching each other; for example:

> *Assistance and simplification of our work will be possible and will be welcome, freeing us from mechanical and boring jobs and preserving human knowledge in order to used it in an optimal way at crucial times.* [Participant 199]

Many comments specified *how* future technology could facilitate the work activity of psychiatrists. Different aspects of the profession were discussed, and a major theme was the role of technology in improving administrative tasks, especially documentation: some respondents couched this as the only benefit to be accrued to psychiatric practice; for example:

> *Only benefit would be with some documentation or ordering.* [Participant 744]

> *The main benefit would be reducing paperwork to psychiatrists.* [Participant 624]
>
> *They would facilitate the bureaucratic work.* [Participant 29]
>
> *Not having to hunch over keyboards would be lovely.* [Participant 91]

However, not all participants agreed: a few believed that technology would lead to "*more bureaucracy*" and "an increase in "*administrative work*"; for example:

> *AI represents bureaucracy and administrators, and we all hate them.* [Participant 47]
>
> *Work intensification through an increased workload.* [Participant 210]

A few psychiatrists anticipated that future technology might play an important role in data-gathering, however comments were typically truncated; for example:

> *Few benefits except for data collection.* [Participant 126]
>
> *Data collection task is easier = benefit.* [Participant 284]
>
> *Better information, data-collection.* [Participant 92]

Some commented on perceived improvements with patient history-taking and the establishment of standardized tests and questionnaires; for example:

> *Artificial intelligence could initially help to know the history, the anamnesis, and potentially the psychiatric symptoms.* [Participant 239]
>
> *There will probably be avatars capable of doing our work for us, in terms of performing clinical interviews.* [Participant 647]

A related commonly perceived benefit was the provision of greater "*consistency*" or "*standardization*" in the application of evidence-based medicine and in clinical decision-making; for example:

> *AI may help psychiatrists to follow standardized protocols better, or to deviate from these protocols with better reasoning.* [Participant 710]

> *Benefits will be to standardize and minimize inter-psychiatrist variability across diagnoses.* [Participant 577]

Many comments indicated a role for "*big data*", "*algorithms*", and "*data analysis*" in augmenting clinical judgments but responses were limited and typically fell short of explanatory detail; for example:

> *Data mining and deep learning.* [Participant 387]

> *Diagnosing and treatment recommendations will become more algorithm based.* [Participant 396]

> *Thorough data analysis (pattern matching, data mining).* [Participant 381]

> *Tools based on memory and finding correlation in data as well as having similar cases, and making comparisons, and making a suggestion based on predictive statistics.* [Participant 670]

With regard to decisions about treatment course, many respondents stressed that future technology will influence various areas, such as the formulation of the treatment plan, and medication decisions; for example:

> *AI will strongly influence the technique of taking medical histories and be helpful in the selection of the best treatments.* [Participant 291]

> *AI's ability to provide more complete information regarding patients' history and mental status will facilitate better management in terms of pharmacotherapy.*
> [Participant 68]

In contrast, only a minority of physicians suggested that future technology will assist in determining the "*effectiveness of therapy*" [Participant 113]. Similarly, the use of brain imaging, genetic testing, and use of AI in monitoring symptoms received only a small number of comments.

*Limited or negative impact on work activity*

Many responses strongly suggested a risk of "*dependence*" on artificial intelligence in clinical decisions that would be inherently problematic; for example:

> *Risks include (…) overdependence of technology leading to complacency in the field or missing important information.* [Participant 3]

> *I feel a possible drawback would be in becoming too reliant on these advantages and becoming lost if there is a system failure* [Participant 544]

> *Dependence on technology = disadvantage.* [Participant 547]

A minority of comments also suggested that future technology might result in a reduction of psychiatric skills and that psychiatrists may lose their "*critical thinking*"; for example:

> *May lead to less skilled mental health staff.* [Participant 511]

> *Decision-making process will be based on low-quality statistical data, and this is not in patient's interests.* [Participant 238]

Going further, numerous comments were associated with considerable skepticism that future technology might ever replace the "*art of medicine*" and that technology would "*oversimplify*" decisions; for example:

> *Psychiatry is an art. Not a science that you plug in symptoms into an algorithm and pop out a diagnosis and treatment plan and prognosis.* [Participant 581]

> *Medicine is not black-and-white, but it is unlikely that an artificial intelligence will be able to detect that and make appropriate medical decisions on a regular basis without human intervention.* [Participant 44]

More strongly, some psychiatrists surveyed stated that they do not expect future technology to impact the general professional status; for example:

> *Inapplicable to psychiatry.* [Participant 214]

> *Will never work for psychiatry. Maybe benefits to other medical specialties.* [Participant 744]

> *It's not really a concern.* [Participant 676].

> *Hardly anything will change, because many evaluations based on experience depend particularly on the intuition of the physician.* [Participant 123]

> *I do not see psychiatry as changing much, it is one of the few fields that bases diagnoses on face to face interviews, non-verbal language etc.* [Participant 154]

Finally, multiple comments expressed uncertainties about the impact of technology on the status of the profession, with many psychiatrists admitting they were "*unsure*" or "*don't know*".

**Consequences of Future Technology at a Systems Level**

Comments encompassed a number of themes related to the impact of future technology on psychiatry at a systems level. The majority of these responses tended to be optimistic, with comments focusing on greater access to psychiatric care; lower costs; and improved efficiencies.

*Access to care*

Many participants described the many ways that technology could increase access to care particularly in remote or underserviced settings; for example:

> *It will be of great benefit in area where there is shortage of psychiatrists.* [Participant 104]

> *There is an already severe deficit for access to care to psychiatrist and this may bridge the gap.* [Participant 374]

> *It could improve the access to care and continuity of the same.* [Participant 714]

> *Benefits would be that patients can be seen in a timely fashion, as there is a shortage of psychiatrists and treatment facilities.* [Participant 786]

*Costs*

Some psychiatrists speculated that technology could impact the cost of care. Many of these comments mentioned the potential benefit to health care organizations and insurance companies; for example:

> *It will be possible to access treatments at lower cost.* [Participant 685]

> *The benefits are likely purely financial and in favor of large managed care orgs.* [Participant 10]

> *The benefits exist in regards to cost savings to insurers and health care organizations.* [Participant 528]

*Increased efficiencies*

Multiple participants commented on the potential for more efficient provision of care; for example:

> *Efficiency, equity, and universality and sustainability.* [Participant 185]
>
> *Better organization of services and use of resources.* [Participant 699]
>
> *It will improve the relationship with the patient by delegating time dedicated to keeping records to computers.* [Participant 708]
>
> *I feel the potential benefit is to save time in analyzing certain data for trends and quicker analysis of complex histories and faster retrieval of vital information.* [Participant 544]

*Scientific Innovation and Knowledge Translation*

Only a few comments highlighted the potential for technology to stimulate scientific advancement, such as the facilitation of knowledge translation, increased knowledge exchange, or more specifically the identification of new biological markers or neuroimaging techniques:

> *Potential benefits: support in the exploratory, diagnostic and treatment process by considering all clinical variables and having scientific information always up to date.* [Participant 649]
>
> *Being able to obtain the right information on all accumulated advances and experience in psychiatric treatment. (…) Exchange with colleagues about the development in neuro-imaging techniques and description of these by experts at a*

> *distance, making these increasingly affordable and easy to do, as well as at a lower cost.* [Participant 743]

> *As data points increase, with the addition of microbiomes, it will be necessary to have AI there to crunch the data into meaningful and interpretable factors guiding approaches toward wellness.* [Participant 523]

## Discussion

### Principal Findings

This extensive qualitative study provides cross-cultural insight into the views of practicing psychiatrists about the potential influence of future technology on psychiatric care (see Box 1). A dominant perspective was that machines would never be able to replace relational aspects of psychiatric care, including empathy and from developing a therapeutic alliance with patients. For the majority of psychiatrists these facets of care were viewed as essentially human capacities.

Psychiatrists' expressed divergent views about influence of future technology on the status of the profession and the quality of medical care. At one extreme, some psychiatrists considered outright replacement of the profession by AI was likely; yet others believed technology would incur no changes to psychiatric services. Many speculated that AI would fully undertake administrative tasks such as documentation, the vast majority of participants predicted that 'man and machine' would collaborate to undertake key aspects of psychiatric care such as diagnostics and treatment decisions. Participants were split over whether AI would might ultimately reduce medical error, or improve diagnostic and treatment decisions. Although many believed that AI could augment doctors' roles, they were skeptical that technology would ever be able to fully undertake medical decisions without human input. For many participants diagnostics and other clinical decisions were quintessentially human skills. Relatedly, risk of overdependence on technology in driving medical error was a common concern.

More positively, many respondents felt technology would be fairer and less biased than humans in reaching clinical decisions. Similarly, participants expressed optimism that technology would play a key role in undertaking administrative duties, such as documentation. Other expected benefits from future technology included improved access to psychiatric care, reduced costs, and increased efficiencies in healthcare systems.

**Technology and Human Interactions**

Although psychiatrists, like informaticians, were optimistic that technology would increase access to psychiatric care, particularly among underserved populations,[] they were cynical that technological advancements could fully replace the provision of human-mediated empathy and relational aspects of care. These views are also shared by many AI experts who also argue that by outsourcing some aspects of medical care to machine learning, physicians will be freed up to invest more time on humanistic elements of medicine, including the provision of longer, face-to-face doctor-patient interactions.[] Nonetheless, drawing on findings in the nascent field of affective computing, it is worth noting that some informaticians speculate that in the long-term, computers may play a critical role in augmenting or replacing human-mediated empathy; for example, emerging studies suggest that under certain conditions, computers can surpass humans when it comes to accurate detection of facial expressions, and personality profiling.[]

Interestingly, very few psychiatrists discussed telepsychiatry despite its potential to increase patient access and adherence to care, however this may have been due to the emphasis on machine learning and artificial intelligence. Technical quality and issues of privacy and confidentiality remain key drawbacks with this medium (see: *Regulation of mHealth* and *Ethical Issues,* below) but patients report high levels of satisfaction, convenience, and comfort with this approach, and evidence indicates that telepsychiatry provides comparable reliability and clinical outcomes as face-to-face consultations.[] Similarly, despite a growing body of research to support digital cognitive behavioral therapy,[] there was limited discussion among psychiatrists about the role of future technology encroaching on psychological treatments.

**The Scope of AI in Psychiatry**

Responses revealed that psychiatrists have myriad, often disparate views about the value of artificial intelligence on the future of their profession. Notwithstanding the wide spectrum of opinion, similar to the views of many experts, a dominant, overarching theme was speculation about a hybrid collaboration between 'man and machine' in undertaking psychiatric care.[] Like informaticians, in particular, many participants highlighted the potential for AI in risk detection and preventative care.[] More generally, psychiatrists – like informaticians – were optimistic about the benefits of AI in augmenting patient care, ergonomic and human factors remain ongoing issues in the design of technology. Without due attention to "alert fatigue" and clinical workflow, it is unclear whether AI applications will reap their anticipated potential in improving clinical accuracy, and in strengthening healthcare efficiencies. and reducing costs.[]

Although a considerable number of participants conceived of clinical decisions as essentially and ineffably, a human "art", biomedical informaticians argue that the ability to mine large scale health data for patterns in diagnosis and behavior is where machine learning presents unprecedented potential to disrupt diagnostic, prognostic, and treatment precision, yielding insights about hitherto undetected subtypes of diseases.[] Against the promise of pattern detection mediated by machine learning, many informaticians acknowledge that current AI is far from sufficient to fully undertake diagnostic decisions unaided, and significant breakthroughs will be necessary if machines are to avoid pitfalls in reasoning, and demonstrate causal and counterfactual reasoning capacities necessary to reach accurate medical decisions.[] Importantly, however, and in contrast to many of the physicians surveyed who considered clinical reasoning to be, in essence, a necessarily human capacity, leading AI experts assume that one cannot rule out, a priori, the possibility that technology may one day be fully capable of fulfilling these key medical tasks.

**Technology and Data-collection**

Disparities between psychiatrists and AI experts were apparent in respect of some key developments and debates about the use of technologies in mental health. For example, only a minority of psychiatrists discussed – whether positively or negatively – the role of smart phones in data gathering. So far, however, encouraging evidence demonstrates that utilizing customized smart phone apps with patient health questionnaires can help to capture patients' symptoms in real-time, allowing more sensitive diagnostic monitoring.[]

Aside from health information gathered via electronic health records and patient reports, an exponentially increasing volume of data accumulated via *in situ* personal digital devices, especially smartphones usage. Social media posts, apps, purchases, and personal internet history, are already being used to support predictions about patient health, behavior, and wellbeing;[] other passively accumulated data from GPS, accelerometer sensors, text and call logs, and screen on/off time can be used to infer mobility, sociability, and other behaviors of smartphone users. Collectively, so-called 'digital phenotyping' provides a novel, indirect, and nontraditional route to yield inferences about patients' health status; it also presents a novel challenge to orthodox boundaries of traditional medical expertise.[] Scarce reflection on these issues among respondents contrasts with the predictions of biomedical informaticians who argue that apps and mobile technologies will play an increasing role in accumulating salient personal health information. Wearable devices, it is argued, will help to facilitate real-time monitoring of signs and symptoms, improving accuracy and precision in information gathering, and helping to avoid barriers associated with routine check-ups, such as missed appointments, personnel shortages, and costs on mental health services.[]

**Patients' Preferences and Mobile Health**

Some psychiatrists argued that interfacing with technology would not be acceptable to many patients who would prefer to receive care from doctors. Previous survey research in mobile health (mHealth) undermines the certitude of these claims; for example, a study by Boeldt and colleagues found that patients were more comfortable with the use of technology performing diagnostics than physicians.[] Studies also show that patients from diverse socioeconomic and geographical regions express willingness to use apps to support symptom tracking, and illness self-management.[] Evidence also suggests that some patients with schizophrenia already use technology to manage their symptoms, or for help-seeking.[] In one recent US survey of 457 adults identifying with schizophrenia, and schizoaffective disorders, 42% "often" or "very often" reported listening to music or audio files to help block or manage voices; 38% used calendar functions to manage symptoms, or sent alarms or reminders; 25% used technology to develop relationships with other individuals who have a lived experience related to mental illness; and 23% used technology to identify coping strategies.[]

Indeed, previously it was assumed that severity of mental health symptoms would pose a barrier to interest in mHealth;[] however, studies show that patients with serious conditions, including psychosis, indicate high levels of interest in the use of mobile applications to manage and track their symptoms and illness.[] As Torous et al argue, it may be that patients are more comfortable using mobile technology to report and monitor symptoms than earlier methods such as sending text messages to clinicians, and that such a medium reduces stigma.[] Relatedly, the co-production of medical notes – for example, patients entering information via semi-structured online questions prior to medical appointments – may also play a role in reducing barriers to help-seeking.[] Although research is ongoing, initial disclosures of symptoms via online patient portals may mitigate stigmatization and feelings of embarrassment in initiating conversations about mental health issues with physicians.

Despite patient interest and evidence of high adoption rates for health and wellness apps, there remain well documented problems with drop off rates, and how to design for continuance – issues that surveyed psychiatrists did not directly discuss.[]

**Regulation of mHealth**

Conspicuously, participants provided scarce commentary about the regulatory ramifications of artificial intelligence on patient care. An estimated excess of 10,000 apps related to mental health are now available for download; the vast majority of these apps have not been subject to randomized controlled trials (RCTs), and many may even provide harmful 'guidance' to users.[] While recent meta-analyses and systematic reviews indicate that a number of safe, evidence-based apps exist for monitoring symptoms of depression, and schizophrenia, and for reducing symptoms of anxiety, patients and clinicians lack adequate guidelines to facilitate recommendations.[]

On the other hand, many psychiatrists expressed enthusiasm about the potential of future technology to provide more objective, and less biased clinical judgments. This optimism appeared to overlook concerns associated with "algorithmic biases" – the risk of discrimination against patients, associated with inferior design and implementation of machine learning algorithms.[] As AI experts and ethicists warn, bias can become baked into algorithms when demographic groups (for example, along the lines of ethnicity, gender, or age) are underrepresented in training phases of machine learning. Without adequate regulatory standards, in the design and ongoing evaluation of algorithms, medical decisions informed by machine learning may exacerbate rather than diminish discrimination arising in clinical contexts.

The US Food and Drug Administration (FDA) has so far adopted a deliberately cautious approach to clarifying medical software regulations.[] Some tech companies have emerged as, "default arbiters and agents responsible for releasing (and on some occasions, withdrawing) applications".[] As medical legal experts warn, allowing unregulated market forces to determine 'kitemarks' of medical standards, is inadequate to protect patient health.[]

**Ethical Issues**

Related to regulatory issues, few comments – only nine in total – weighed in on ethical issues related to protections for sensitive personal data. Loss of patient data and privacy remain serious concerns for mobile applications and telepsychiatry. In 2018 the European Union (EU) enacted its 'General Data Protection Regulation' (GDPR) aimed at ensuring citizens have control of their data, and provide consent for the utilization of their sensitive personal information. The US has considerably weaker data privacy rules, and while similar legislation to the GDPR is mooted to come into effect in California in January 2020, no comparable laws have been enacted at a federal level in the USA nor is there legislative enthusiasm to do so. Notably, these issues have prompted much recent media coverage. Given the gravity of ethical issues surrounding adequate oversight for patient data gathering from apps and mobile technologies, including how they might impact doctor-patient relationships and adequate patient care, and the media coverage that these issues have prompted, it was conspicuous that privacy and confidentiality considerations, received scarce commentary from surveyed psychiatrists.[] Similarly, while may psychiatrists believed future technology would be a boon to patient access, issues of justice related to the 'digital divide' – between those who have ready access to the internet and mobile devices, and those who did not – received no attention.

**Strengths and Limitations**

To our knowledge this is the first survey to undertake qualitative research aimed at understanding psychiatrists' views about how AI/ML will impact their profession. The themes support and expand on findings of an earlier quantitative survey by providing a more refined perspective of psychiatrists' opinions about AI and the future of their profession. Utilizing the Sermo platform enabled us to gain rapid responses from verified and licensed physicians from across the world, and this survey benefits from a relatively large sample size of participants working in different countries across a broad spectrum of practice settings. The diversity of respondents combined with the unusually high response rate for questions requesting comments, are major strengths of the survey.

The study has a number of limitations. Comments were often brief, and because of the restrictions of online surveys it was not possible to obtain a more nuanced understanding of participants' views. Therefore, although a rich and diverse range of opinions was gathered, further qualitative work is warranted to obtain more fine-grained analysis of physicians' opinions about the impact of AI/ML on the practice of psychiatry and on patient care. Furthermore, we did not gather information on physicians' level of knowledge or exposure to the topic or AI/ML in medicine, limiting inferences about awareness, and the depth of participants' reflections.

Notably, some participants explicitly expressed uncertainty about whether AI could benefit medical judgment with some admitting they had limited familiarity with the field.

In-depth qualitative interviews, or focus groups would help to facilitate deeper analysis of psychiatrists' perspectives and their understanding of these fields and their impact on psychiatry. We further recommend that follow-up qualitative work related to these themes should focus on the views of other mental health professionals, including mental health nurse practitioners, psychiatric nurses, clinical psychologists and psychotherapists, and other counselors. Finally, there can be no complete account of the anticipated benefits, harms, and hopes for AI in psychiatry without the ongoing input of mental health patients, and their caregivers.

**Conclusions**

This study provides a foundational exploration of psychiatrists' views about the future of their profession. Perceived benefits and limitations of future technology in psychiatric practice, and the future status of the profession, have been elucidated. A variety of perspectives were expressed reflecting a wide range of opinions. Overwhelmingly, participants were skeptical about the role of technology in providing empathetic care in patient interactions. Although some participants expressed anxiety about the future of their job, viewing technology as a threat to the status of their profession, the dominant perspective was a prediction that human medics and future technology would work together. However, participants were divided over whether this collaboration might ultimately improve or harm clinical decisions including diagnostics and treatment recommendations, and overreliance on machine learning was a recurrent theme. Similar to biomedical informaticians, participants were also hopeful that technology might improve care at a systems level, improving access, increasing efficiencies, and lowering healthcare costs.

While psychiatrists' opinions often mirrored the predictions of AI experts, results also revealed worrying omissions in respondents' comments. In light of high levels of patient interest in mental health apps, the effectiveness, reliability, and safety of machine learning technologies present serious ethical, legal, and regulatory considerations that require the sustained engagement of the psychiatric community.[] So far, the efficacy and safety of the overwhelming majority of downloadable mHealth apps have yet to be demonstrated.[] Moreover, in contrast to the views of many leading informaticians, psychiatrists were often enthusiastic that technology would reduce biases in decision-making; however, without further regulatory attention to standards of design within machine learning, it is unclear that algorithms will help to redress rather than deepen healthcare disparities. Against these considerations, steadfast leadership is required from the psychiatric community to help patients navigate mobile health apps, and to advocate for guidelines with respect to digital tool, to ensure current mHealth as well as emerging technologies, do not jeopardize standards of safety and trust in patient care.

Finally, given the sheer breadth of opinion, and oversights,[] it is conceivable that many practitioners are disengaged from the literature on healthcare AI.[] Recent physician surveys suggest medical education on health technology "leaves much room for improvement".[] For example, an extensive cross-sectional survey of EU medical schools found that fewer than a third (90/302, 30%) offered any kind of health information technology training as part of medical degree courses. Similarly, a recent survey of physicians in South Korea reported that only 6% (40/669) of those surveyed described "good familiarity with AI".[] While gaps in knowledge are understandable given the volume of medical course curricula and the time pressures of clinical practice, we conclude that the medical community must do more to raise awareness of AI among current and future physicians. Lacking adequate education about machine learning technology and its potential to impact the lives of patients, psychiatrists will be ill-equipped to steer mental health care in the right direction.

## Declarations

Conflicting interests: Dr Doraiswamy has received research grants from and/or served as an advisor or board member to government agencies, technology and healthcare businesses, and advocacy groups for other projects in this field.

Funding: This survey was supported by Sermo which provided the platform for the survey and funded the physicians. Dr Blease was supported by an Irish Research Council- Marie Skłodowska-Curie Fellowship. Dr Locher was funded by a Swiss National Science Foundation grant (P400PS_180730).

Ethical approval: This study was deemed exempt research by Duke University Institutional Review Board.

Guarantor: CB is the guarantor of this article.

Contributorship: Conceived & initiated project: CB, PMD. Analyzed results: CB, CL, MLC. Wrote first draft: CB. Contributed to revisions: CB, CL, MLC, PMD.

Acknowledgements: The authors would like to than Sermo, especially Peter Kirk and Joanna Molke, for their collaboration. We would also like to express our gratitude to the doctors who participated in this survey and shared their valuable insights.

43. Yu K-H, Kohane IS. Framing the challenges of artificial intelligence in medicine. BMJ Qual Saf. 2018;bmjqs–2018.

44. Singh H, Spitzmueller C, Petersen NJ, Sawhney MK, Sittig DF. Information overload and missed test results in electronic health record–based settings. JAMA internal medicine. 2013;173(8):702–704.

45. Murdoch TB, Detsky AS. The inevitable application of big data to health care. Jama. 2013;309(13):1351–1352.

46. Yu K-H, Beam AL, Kohane IS. Artificial intelligence in healthcare. Nature Biomedical Engineering. 2018;2(10):719.

47. Torous J, Staples P, Shanahan M, Lin C, Peck P, Keshavan M, et al. Utilizing a personal smartphone custom app to assess the patient health questionnaire-9 (PHQ-9) depressive symptoms in patients with major depressive disorder. JMIR mental health. 2015;2(1):e8.

48. Torous J, Staples P, Onnela J-P. Realizing the potential of mobile mental health: new methods for new data in psychiatry. Current psychiatry reports. 2015;17(8):61.

49. Insel TR. Digital phenotyping: technology for a new science of behavior. Jama. 2017;318(13):1215–1216.

50. Onnela J-P, Rauch SL. Harnessing smartphone-based digital phenotyping to enhance behavioral and mental health. Neuropsychopharmacology. 2016;41(7):1691.

51. Fogel AL, Kvedar JC. Artificial intelligence powers digital medicine. NPJ Digital Medicine. 2018;1(1):5.

52. Cui M, Wu X, Mao J, Wang X, Nie M. T2DM self-management via smartphone applications: a systematic review and meta-analysis. PLoS One. 2016;11(11):e0166718.

53. Firth J, Cotter J, Torous J, Bucci S, Firth JA, Yung AR. Mobile phone ownership and endorsement of "mHealth" among people with psychosis: a meta-analysis of cross-sectional studies. Schizophrenia bulletin. 2015;42(2):448–455.

54. Gay K, Torous J, Joseph A, Pandya A, Duckworth K. Digital technology use among individuals with schizophrenia: results of an online survey. JMIR mental health. 2016;3(2):e15.

55. Ennis L, Rose D, Denis M, Pandit N, Wykes T. Can't surf, won't surf: the digital divide in mental health. Journal of Mental Health. 2012;21(4):395–403.

56. Torous J, Friedman R, Keshavan M. Smartphone ownership and interest in mobile applications to monitor symptoms of mental health conditions. JMIR mHealth and uHealth. 2014;2(1):e2.

57. Mafi JN, Gerard M, Chimowitz H, Anselmo M, Delbanco T, Walker J. Patients contributing to their doctors' notes: insights from expert interviews. Annals of internal medicine. 2018;
39

## Table 1. Open Comment Questions embedded in Survey

| |
|---|
| 1. Please briefly describe the way(s) you believe artificial intelligence/machine learning will change psychiatrists' jobs in the next 25 years.* |
| 2. Please provide any brief comments you may have about the potential benefits and/or potential harms of artificial intelligence/machine learning in psychiatry. |
| 3. We value your opinion. If you have any other comments about this survey topic or recommendations for other questions we should include, please add them below. |

**\* All participants were requested to respond to Questions 2 and 3. However, Question 1 was preceded by the following question:** "*In 25 years, of the following options, in your opinion what is the likely impact of artificial intelligence/machine learning on the work of psychiatrists*". Options included "*No influence (jobs will remain unchanged)*", "*Minimal influence (jobs will change slightly)*"; "*Moderate influence (jobs will change substantially)*" or "*Extreme influence (jobs will become obsolete)*". Participants who selected the first response ["*No influence (jobs will remain unchanged)*"] were not invited to respond to Question 1.

## Table 2: Respondent Characteristics

| Characteristic | Psychiatrists (n=791) *Percentage* |
|---|---|
| **Gender** | |
| Male | 69.5 |
| Female | 29.2 |
| Other | 0.1 |
| Prefer not to say | 1.1 |
| | |
| **Age** | |
| 25-34 | 9.7 |
| 35-44 | 29.3 |
| 45-54 | 26.7 |
| 55-64 | 24.7 |
| 65 and over | 9.6 |
| | |
| **Race/ethnicity** | |
| Asian | 17.6 |
| Black/African/Caribbean | 2.0 |
| Mixed/Multiple ethnic groups | 3.7 |
| White | 64.3 |
| Other ethnic group not listed | 3.2 |
| Prefer not to say | 9.3 |
| | |
| **Practice type** | |
| Private practice | 35.0 |
| Public clinic | 52.0 |
| Academia | 13.0 |
| | |
| **Country where psychiatrist practices** | |
| United States | 34.9 |
| France | 9.7 |
| Italy | 9.4 |
| Germany | 7.5 |
| Spain | 7.2 |
| United Kingdom | 6.3 |
| Russian Federation | 3.8 |
| Australia | 3.2 |
| Japan | 2.8 |
| Mexico | 2.5 |
| Canada | 2.3 |
| Greece | 1.9 |
| China | 1.8 |
| Brazil | 1.5 |
| Turkey | 1.4 |
| Netherlands | 1.0 |
| Belgium | 0.5 |
| Switzerland | 0.4 |
| Norway | 0.3 |
| Portugal | 0.3 |
| India | 0.1 |

## Box 1: Key questions and findings

**What is already known about this topic?**

- Informaticians and experts in artificial intelligence (AI) argue that big data and machine learning (ML) have the potential to revolutionize how psychiatric care is delivered.
- Recent survey evidence suggests that psychiatric patients, including those suffering from severe mental illness express an interest in using mobile technologies to monitor and manage their condition(s).
- To date, in excess of 10,000 apps related to mental health are available to download; the vast majority have not been subject to RCTs.
- Indirectly, data accumulated from *in situ* personal digital devices can also be used to support predictions about patient health, behavior, and wellbeing – this is known as 'digital phenotyping'.

**What are the new findings?**

- 791 psychiatrists from 22 countries responded to an online survey via the physician social networking platform Sermo; 70% were male; 61% were aged 45 or older.
- Overwhelmingly, psychiatrists were skeptical that machines could replace humans in the delivery of empathic care, and in forging therapeutic alliances with patients.
- Many predicted that in the future 'man and machine' would increasingly collaborate on key aspects of psychiatric care, such as diagnostics and treatment decisions; psychiatrists were divided over whether technology would augment or diminish the quality of medical decisions and patient care.
- In contrast to concerns of AI experts, psychiatrists provided limited or no reflection about issues relating to digital phenotyping, and regulatory and ethical considerations related to mobile health.